\documentclass[preprint,12pt]{elsarticle}

\usepackage{amssymb}

\newcommand{\upa}{\uparrow}
\newcommand{\dna}{\downarrow}
\newcommand{\dgr}[1]{#1^{\dagger}}

\journal{Physica C}

\begin{document}

\begin{frontmatter}



\title{Role of charge fluctuation in Q1D organic superconductor ${(TMTSF)_2ClO_4}$}


\author{Yusuke Mizuno$^1$, Akito Kobayashi$^{1,2}$, and Yoshikazu Suzumura$^1$}

\address{$^1$Department of Physics, Nagoya University, Furo-cho, Chikusa-ku, Nagoya, 464-8602 Japan 
 \\ $^2$Institute for Advanced Research, Nagoya University, Furo-cho, Chikusa-ku, Nagoya, 464-8602 Japan}

\begin{abstract}
We have theoretically investigated the spin and charge fluctuations
in the quasi-one dimensional organic superconductor ${\rm (TMTSF)_2ClO_4}$.
Using the extended multi-site Hubbard model,
which contains  four sites
 in a unit cell and
 the transfer energies obtained  by the extended H\"uckel method,
we calculate the linearized gap equation with the random phase approximation, 
to find 
 novel order parameters of  superconductivity 
 due to several kinds of charge fluctuations 
   induced by  the anisotropic intersite repulsive interactions. 
 For the singlet state,
 the order parameter with line nodes appears 
   in the case of  the strong charge fluctuation,  
 while  the order parameter  with anisotropic gap 
  suggested by Shimahara is reproduced  
 in the  spin fluctuation.
The triplet state is also obtained  for the wide parameter range 
   of repulsive interactions 
 due to 
 a cooperation between charge and spin fluctuations. 
\end{abstract}


\end{frontmatter}


\section{Introduction}

In quasi-one dimensional organic conductors TMTSF (tetrametyltetraselenafulvalence) salts, there are 
 several experimental evidences of superconductivity, 
\cite{Bechgaard,taki,jero,shina,lee,lee2,lee3,yone,yone2,yone3}
  which exhibit the large critical field being  much larger than  the Pauli limit.\cite{lee,yone,yone2,yone3}
  For  ${\rm (TMTSF)_2PF_6}$,
   the triplet superconducting(SC) state has been maintained 
  from the measurement of  the Knight shift and the NMR relaxation rate.
\cite{lee2,lee3}
 For ${\rm (TMTSF)_2ClO_4}$, \cite{Bechgaard}
 in the weak magnetic field, 
the singlet state with line nodes has been suggested
   by the 
the Knight shift and relaxation rate of the NMR experiment in the SC phase.\cite{taki,shina} 
Further, it has been suggested that the triplet or FFLO state emerges 
under strong magnetic field.\cite{shina,yone,yone2,yone3}

The specific property of ${\rm (TMTSF)_2ClO_4}$ emerges due to 
   the anion ordering below 24K.
The folded  Fermi surfaces with 
 a unit cell containing four TMTSF molecules
 is obtained from the transfer integrals, which are estimated from 
  the extended H\"uckel method based on 
    the X-ray diffraction measurement.\cite{yoshi,pou,band}

Several theoretical works have been performed to understand the superconductivity 
in the quasi-one dimensional organic conductors.
The single band models for ${\rm (TMTSF)_2PF_6}$
 has been investigated by perturbation theories.
 \cite{tanaka,kuroki,aizawa,aizawa2,bel,kinokon}
 The magnetic field favors the triplet or FFLO superconductivity, \cite{aizawa,aizawa2,bel}
while the on-site repulsion induces the singlet superconductivity in the absence of magnetic field. \cite{kinokon}
The interplay between the spin and charge fluctuations in the triplet superconductivity 
has been investigated by the renormalization group theories.
\cite{duprat,fuseya,nick}
The inter-site repulsions play important roles 
for the triplet superconductivity. \cite{tanaka,kuroki,fuseya,nick}
 In the multi band models for ${\rm (TMTSF)_2ClO_4}$\cite{shima,hase}, 
on the other hand, the singlet superconductivity with anisotropic gap (but without line nodes) 
has been suggested in the absence of magnetic field.\cite{shima}

In the present study, the possible singlet state with line nodes in the absence of magnetic field is investigated, 
since the experiment indicates the existence of line nodes.\cite{yone}
We use the extended multi-site Hubbard model representing the system of ${\rm (TMTSF)_2ClO_4}$,
when the transfer energies are given by the extended H\"uckel method with the X-ray diffraction measurement.\cite{band} 
The unit cell consists of four TMTSF sites owing to the presence of the anion ordering.
Fruther, the roles of the charge fluctuation for the triplet state are also investigated 
on such a multi-site system.
The spin and charge fluctuations are treated by the random phase approximation (RPA) on the site-representation, 
\cite{koba} and the order parameters of the superconductivity are evaluated using the linearized gap equation 
as described in \S 2. 
The singlet and triplet superconductivities are investigated 
by varying the magnitude of the anisotropic nearest-neighbor repulsive interactions
      with the fixed on-site interaction in \S 3.
The origins and properties of those SC states are discussed in \S 4.
In \S 5, summary and discussion are given.

\section{Formulation}
\subsection{Model}

\begin{figure}[t]
 \begin{center}
 \includegraphics*[scale=0.35,clip]{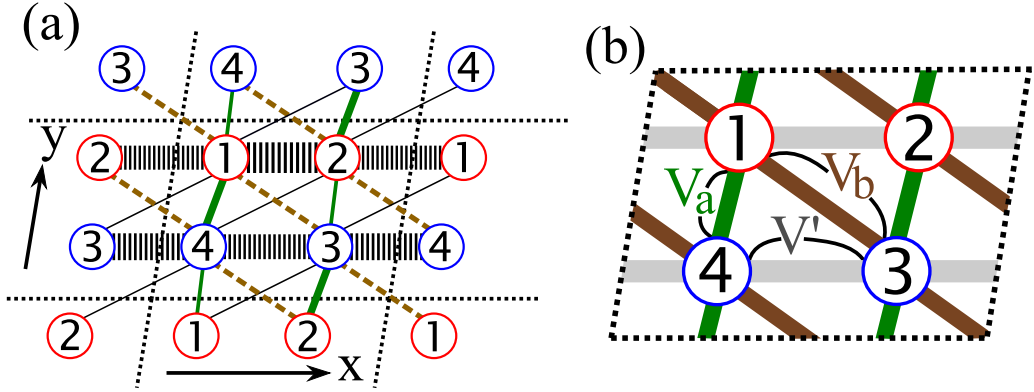}
 \caption{
(a) A model describing the two-dimensional electronic system 
for ${\rm (TMTSF)_2ClO_4}$.
The x-axis is taken as the stacking direction.
The unit cell (the dotted line) 
  consists of 4 molecules with several transfer energies 
 $t_{{\bf d} ; \alpha \beta}$.
(b) Inter-site repulsive interactions $V_a,V_b$ and $V'$, 
where $V'$ is discarded.
}
 \label{fig:hopping}
 \end{center}
\end{figure}

Figure \ref{fig:hopping} (a) shows the transfer integrals describing the two-dimensional (2D) electronic system 
for ${\rm (TMTSF)_2ClO_4}$, where we choose the x-axis along the stacking direction 
 of TMTSF molecules.
The thickness of the lines connecting between the molecules represents 
 the relative magnitude of $t_{{\bf d} ; \alpha \beta}$.
Figure \ref{fig:hopping} (b) displays the inter-site replusive interactions,
which are taken in the present model.

Based on Figs. \ref{fig:hopping} (a) and (b),
the extended multi-site Hubbard model is given by
\begin{equation}
{\cal H} = {\cal H}_0 + {\cal H}',
\end{equation}
with
\begin{equation}\label{eq:h-kinetic2}
{\cal H}_0 = \sum_{{\bf i} {\bf d} \alpha \beta \sigma} ((t_{{\bf d} ; \alpha \beta} - \mu \delta_{\alpha ,\beta} \delta({\bf d})) \dgr{a_{{\bf i} \alpha \sigma}} a_{{\bf i} + {\bf d} \beta \sigma} + h.c.),
\end{equation}
and 

\begin{equation}\label{eq:h-interaction2}
{\cal H}' = \sum_{{\bf i} \alpha} U \dgr{a_{{\bf i} \alpha \upa}} \dgr{a_{{\bf i} \alpha \dna}} a_{{\bf i} \alpha \dna} a_{{\bf i} \alpha \upa}
 + \sum_{{\bf i} {\bf d} \alpha \beta \sigma \sigma'} V_{{\bf d};\alpha \beta} \dgr{a_{{\bf i} \alpha \sigma}} \dgr{a_{{\bf i}+{\bf d} \beta \sigma'}} a_{{\bf i}+{\bf d} \beta \sigma'} a_{{\bf i} \alpha \sigma},
\end{equation}
where  
  ${\bf i}(=(1,1) \cdots (\sqrt{N_L},\sqrt{N_L}))$ denotes a two dimensional lattice vector  for the unit cell,
 and ${\bf d}$ is a vector connecting 
   different cells.
   Quantities $\alpha$ and $\beta(=1,2,3,4)$ are indices of molecules in the unit cell and
 $\sigma(=\upa,\dna)$ represent spin. 
The transfer energies, $t_{{\bf d} ; \alpha \beta}$, are obtained by the extended H\"uckel method based on 
    the X-ray diffraction measurement.\cite{pou,band,yoshi}
Coupling constants $U$ and $V_{{\bf d};\alpha \beta}$ correspond to the on-site and nearest neighbor Coulomb interactions, respectively.
For $V_{{\bf d};\alpha \beta}$, 
 there is a relation,
 $V_{{\bf d};\alpha \beta}=V_{{\bf d};\beta \alpha}=V_{- {\bf d};\alpha \beta}$.
The unit of the energy is taken as eV.

We investigate the superconductivity by varying the nearest neighbor Coulomb interactions between the TMTSF chains, i.e., 
$V_a$ along the $y$ axis and $V_b$ along the $x+y$ axis (see Fig. \ref{fig:hopping} (b)).
In the RPA, the critical value of the nearest neighbor Coulomb interaction along the intrachain, $V'$, 
at which the charge fluctuation diverges, is more than twice of those of $V_a$ and $V_b$.
This comes from the fact that the role of the nesting vector is incompatible with that of $V'$.
The same fact has been shown in the renormalization group theory.\cite{nick}
Thus $V'$ is discarded in the present study.

Using Fourier transformation, the Hamiltonian is rewritten as
\begin{eqnarray}
{\cal H} &=& \sum_{{\bf k} \sigma \alpha \beta} \epsilon_{\alpha \beta}({\bf k}) \dgr{a_{{\bf k} \sigma \alpha}} a_{{\bf k} \sigma \beta} \nonumber \\
 &+& \frac{1}{N}\sum_{{\bf k} {\bf k'} {\bf q} \alpha} U \dgr{a_{{\bf k} + {\bf q} \upa \alpha}} \dgr{a_{{\bf k'} - {\bf q} \dna \alpha}} a_{{\bf k'} \dna \alpha} a_{{\bf k} \upa \alpha} \nonumber \\
 &+& \frac{1}{2N} \sum_{{\bf k} {\bf k'} {\bf q} \sigma \sigma' \alpha \beta} V_{\alpha \beta}({\bf q})
 \dgr{a_{{\bf k} + {\bf q} \sigma \alpha}} \dgr{a_{{\bf k'} - {\bf q} \sigma' \beta}} a_{{\bf k'} \sigma' \beta} a_{{\bf k} \sigma \alpha},
\end{eqnarray}
where 
\begin{eqnarray}
\epsilon_{12}({\bf k}) &=& t_{12} + t'_{12} e^{i k_x} ,\nonumber \\
\epsilon_{13}({\bf k}) &=& t_{13} + t'_{13} (e^{i k_x} + e^{-i k_y}) + t''_{13} e^{i (k_x - k_y)}, \nonumber \\
\epsilon_{14}({\bf k}) &=& t_{14} + t'_{14} e^{-i k_y}, \nonumber \\
\epsilon_{23}({\bf k}) &=& t_{23} + t'_{23} e^{-i k_y}, \nonumber \\
\epsilon_{24}({\bf k}) &=& t_{24} + t'_{24} (e^{-i k_x} + e^{-i k_y}) + t''_{24} e^{-i (k_x + k_y)}, \nonumber \\
\epsilon_{34}({\bf k}) &=& t_{34} + t'_{34} e^{-i k_x},
\end{eqnarray}
\begin{eqnarray}
V_{13}({\bf q}) &=& V_{13} + V'_{13} (e^{i q_x} + e^{-i q_y}) + V''_{13} e^{i (q_x - q_y)}, \nonumber \\
V_{14}({\bf q}) &=& V_{14} + V'_{14} e^{-i q_y}, \nonumber \\
V_{23}({\bf q}) &=& V_{23} + V'_{23} e^{-i q_y}, \nonumber \\
V_{24}({\bf q}) &=& V_{24} + V'_{24} (e^{-i q_x} + e^{-i q_y}) + V''_{24} e^{-i (q_x + q_y)},
\end{eqnarray}
with $\epsilon_{\beta \alpha} = \epsilon^{\ast}_{\alpha \beta},V_{\beta \alpha} = V^{\ast}_{\alpha \beta}$, 
and the other elements are zero.
The transfer energies are given by\cite{pou,band,yoshi} 
$t_{12}=0.413, t'_{12}=0.324,
t_{34}=0.335$ and $t'_{34}=0.362$ along conduction chains,
$t_{23}=t'_{14}=-0.050, 
t_{14}=t'_{23}=-0.100$
 along y-axis,
and 
$t'_{24} = 0.070, 
t_{24}=t"_{24}=0.020, t_{13}=t"_{13}=0.071, 
 t'_{13}=0.021$ 
where $t$, $t'$ and $t''$ correspond to transfer energies between molecules 
 in a unit cell, those between the nearest neighbor cells, and those between the next-nearest neighbor cells, respectively. 
 Note $t_{\alpha \beta}=t_{\beta \alpha}$.
 We choose 
  $V_{14}=V_{23}=V'_{14}=V'_{23}=V_a$ for nearest neighbor, i,e, 
 ${\bf d} =(\pm 1, 0), (0, \pm 1) $
 and $V_{13}=V'_{24}=V'_{24}=V''_{13}=V_b$
 for next nearest neighbor, i.e., 
 ${\bf d} =(\pm 1, \pm 1), (\pm 1, \mp 1)$.
(See Fig. \ref{fig:hopping}(b).)
The tight-binding Hamiltonian Eq. (\ref{eq:h-kinetic2}) is diagonalized by
\begin{equation}\label{eq:diago}
\sum_{\beta}^4 (\epsilon_{\alpha \beta}({\bf k}) - \mu \delta_{\alpha \beta}) d_{\beta}^{A}({\bf k}) = \xi_{{\bf k}}^{A} d_{\alpha}^{A}({\bf k}),
\end{equation}
where $A (=1,2,3,4)$ is an index of the band representation 
 and $\xi^{A}({\bf k})$ denotes the eigen energy 
  measured from the Fermi energy ($= \mu$).
note that the eigenvector $d_{\beta}^{A}({\bf k})$ gives 
 the transformation from 
  the  site representation to the  band representation.

\begin{figure}[tb]
 \begin{center}
 \includegraphics*[scale=0.55,clip]{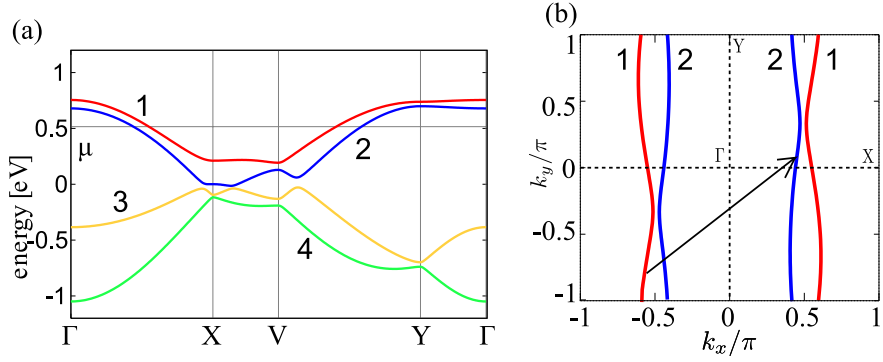}
  \caption{
 (a) Energy Band for ${\rm (TMTSF)_2ClO_4}$ given by Eq. (\ref{eq:h-kinetic2}) with the 3/4 filling,
   where the two bands 1 and 2 from the top 
    cross the Fermi energy, $\mu$.
 (b)
 The Fermi surfaces for the first and second bands,
 where
 the arrow represents the nesting vector. }
 \label{fig:bands}
 \end{center}
\end{figure}

\begin{figure}[tb]
 \begin{center}
 \includegraphics*[scale=0.75,clip]{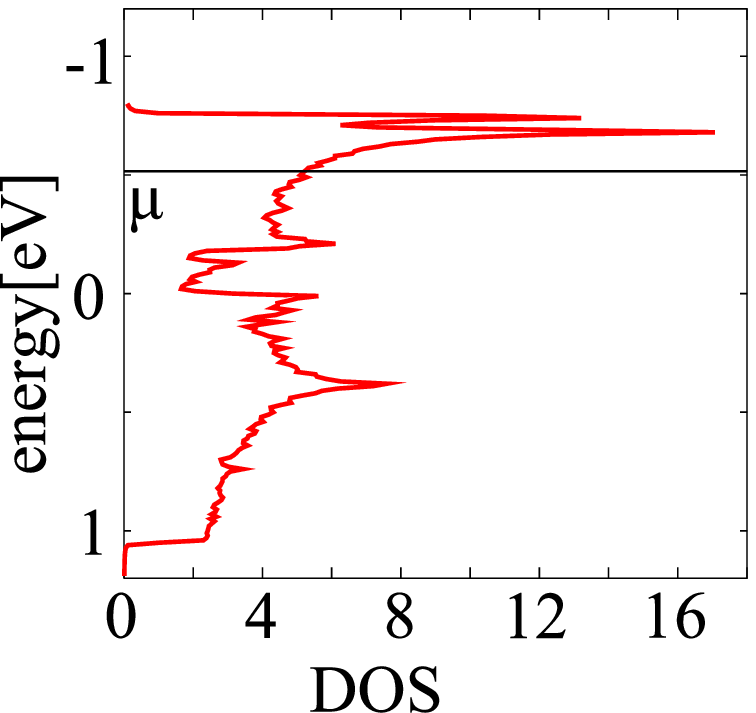}
  \caption{
  The density of states (horizontal axis).}
 \label{fig:fermis}
 \end{center}
\end{figure}

In Fig. \ref{fig:bands}(a), the energy band spectrum is shown.
There are four bands due to four molecules in the unit cell 
where the first and second bands from the top crosses the Fermi surface.
 The shape of quasi one-dimensional Fermi surface is shown in Fig. \ref{fig:bands}(b) representing the folded Fermi surface.
In Fig. \ref{fig:fermis}, the density of states ,$D(\epsilon)$, is shown where
\begin{equation}
D(\epsilon) = -  \frac{1}{N \pi}  \sum_{{\bf k} \sigma A} {\rm Im} \left( \frac{1}{\epsilon + i \delta - \xi^{A}_{{\bf k}} + \mu} \right), 
\end{equation}
and sum rule is given by $\int D(\epsilon) d \epsilon = 8$.
These double one-dimensional Fermi surfaces have been observed by AMRO\cite{yoshi}.

\subsection{Pairing Interactions and Gap Equation}

The spin and charge fluctuations in the weak coupling regime are treated by RPA.
\cite{tanaka,kuroki,aizawa,aizawa2,kinokon,koba2,koba}
where the site representation\cite{koba2} 
 for singlet and triplet pairing interactions 
 are given respectively as
\begin{equation}\label{eq:p-singlet}
\hat{P}^{\rm Singlet} = \hat{U} + \hat{V} + \frac{3}{2} \hat{U} \hat{\chi}^S \hat{U} -\frac{1}{2}(\hat{U} + 2 \hat{V}) \hat{\chi}^C (\hat{U} + 2 \hat{V}),
\end{equation}
\begin{equation}\label{eq:p-triplet}
\hat{P}^{\rm Triplet} = \hat{V} - \frac{1}{2} \hat{U} \hat{\chi}^S \hat{U} -\frac{1}{2}(\hat{U} + 2 \hat{V}) \hat{\chi}^C (\hat{U} + 2 \hat{V}).
\end{equation}
The hat denotes a matrix representation
 with an element $(\hat{f})_{\alpha \beta}=f_{\alpha \beta}$.
The pairing interactions are obtained from 
  $\hat{\chi}^S$ and $\hat{\chi}^C$ which denote 
 spin susceptibility and charge susceptibility;
\begin{equation}\label{eq:xs}
\hat{\chi}^S = (\hat{I} - \hat{\chi}^0 \hat{U} )^{-1} \hat{\chi}^0,
\end{equation}
and 
\begin{equation}\label{eq:xc}
\hat{\chi}^C = (\hat{I} + 2 \hat{\chi}^0 \hat{V} + \hat{\chi}^0 \hat{U} )^{-1} \hat{\chi}^0.
\end{equation}

The quantity $\hat{\chi}^0$ is an irreducible susceptibility which is given by
\begin{eqnarray}\label{eq:x0}
(\hat{\chi}^0({\bf q}))_{\alpha \beta} &=& \frac{-1}{N_L} \sum_{A B {\bf k}}
d_{\alpha}^{A}({\bf k}+{\bf q}) d_{\beta}^{\ast A}({\bf k}+{\bf q}) d_{\alpha}^{\ast B}({\bf k}) d_{\beta}^{B}({\bf k}) \nonumber \\
& \times & \frac{f(\xi_{{\bf k} + {\bf q}}^{A}) - f(\xi_{{\bf k}}^{B})}{\xi_{{\bf k} + {\bf q}}^{A} - \xi_{{\bf k}}^{B}},
\end{eqnarray}
where $f(x)=(e^{x/T}+1)^{-1}$ is the  Fermi distribution and $T$ denotes temperature.

In order to examine a symmetry of order parameter of the superconductivity,
 we solve numerically a linearized gap equation
   given by, 
\begin{equation}\label{eq:linear-gapeq}
\lambda \Sigma^a_{\alpha \beta}({\bf k}) = - \frac{T}{N_L} \sum_{{\bf k'} \alpha' \beta'} \left[ (\hat{P}^A({\bf k}-{\bf k'}))_{\alpha \beta}
G^0_{\alpha \alpha'}({\bf k'}) G^0_{\beta \beta'}({\bf -k'})  \right]
 \Sigma^a_{\alpha \beta}({\bf k'}) . 
\end{equation}
$\Sigma^a_{\alpha \beta}({\bf k})$ is anomalous self energy and 
$G^0_{\alpha \alpha'}({\bf k'})$ denotes a bare Green's function
 written as 
\begin{equation}\label{eq:green-0}
G^0_{\alpha \beta}({\bf k},i \varepsilon_n) = \sum_{A} d_{\alpha}^{A}({\bf k}) d_{\beta}^{\ast A}({\bf k}) \frac{1}{i \varepsilon_n - \xi^{A}_{{\bf k}}},
\end{equation}
where $\varepsilon_n=(2n+1)\pi T$ is Matsubara-frequency ($k_B=1$) and n is an integer.
The quantity $\hat{P}^A$ (A = Singlet, Triplet) denotes 
  the paring interactions for singlet state
  and triplet state which are 
 rewritten as 
\begin{equation}\label{eq:psinglet2}
\hat{P}^{\rm Singlet} = \hat{P}^{S_S} + \hat{P}^{C},
\end{equation}
and 
\begin{equation} \label{eq:ptriplet2}
\hat{P}^{\rm Triplet} = \hat{P}^{S_T} + \hat{P}^{C}.
\end{equation}
The r.h.s.  of Eqs. (\ref{eq:psinglet2}) and (\ref{eq:ptriplet2}) is given by
\begin{equation}\label{eq:pss}
\hat{P}^{S_S} = \hat{U} + \frac{3}{2} \hat{U} \hat{\chi}^S \hat{U},
\end{equation}
\begin{equation}\label{eq:pst}
\hat{P}^{S_T} = - \frac{1}{2} \hat{U} \hat{\chi}^S \hat{U},
\end{equation}
\begin{equation}\label{eq:pc}
\hat{P}^{C} = \hat{V} -\frac{1}{2}(\hat{U} + 2 \hat{V}) \hat{\chi}^C (\hat{U} + 2 \hat{V}).
\end{equation}
 
Note that $\hat{P}^{S_S}$ and $\hat{P}^{S_T}$
 represent spin fluctuation, and $\hat{P}^{C}$ denotes the charge fluctuation.

The transition temperature for the SC state, $T_{\rm c}$, is determined 
by the condition, $\lambda =1$.  
The gap function at $T=0$, $\Delta_{\alpha \beta}({\bf k})$, is estimated from 
 $\Delta_{\alpha \beta}({\bf k}) = g \Sigma^a_{\alpha \beta}({\bf k})$,
 where $g$ is chosen to reproduce the gap which is expected from $T_{\rm c}$. \cite{koba2}
Using $\Delta_{\alpha \beta}({\bf k})$, 
we get the quasi-particle bands $E_{{\bf k}}^{A'}$ 
by diagonalizing Hamiltonian ${\cal H}$,
\begin{eqnarray}\label{eq:bcsH}
{\cal H} &=& \sum_{{\bf k} \alpha \beta \sigma} \epsilon_{\alpha \beta}({\bf k}) \dgr{a_{{\bf k} \alpha \sigma}} a_{{\bf k} \beta \sigma} \nonumber \\
 &-& \sum_{{\bf k} \alpha \beta} \Delta_{\alpha \beta}({\bf k}) \dgr{a_{{\bf k} \alpha \upa}} \dgr{a_{- {\bf k} \beta \dna}} \nonumber \\
 &-& \sum_{{\bf k} \alpha \beta} \Delta^{\ast}_{\alpha \beta}({\bf k}) a_{- {\bf k} \beta \dna} a_{{\bf k} \alpha \upa} ,
\end{eqnarray}
\begin{equation}\label{eq:diago2}
\sum_{\beta'}^8 \tilde{\epsilon}_{\alpha' \beta'}({\bf k}) d_{\beta'}^{A'}({\bf k}) = E_{{\bf k}}^{A'} d_{\alpha'}^{A'}({\bf k}),
\end{equation}
where
\begin{eqnarray}\label{eq:nanbu}
\tilde{\epsilon}({\bf k}) =
\left(
\begin{array}{cc}
\hat{\epsilon}({\bf k})- \mu \delta_{\alpha \beta} & - \hat{\Delta}({\bf k}) \\
-\hat{\Delta}^{\dagger}(-{\bf k}) & - ^t \hat{\epsilon}(-{\bf k}) + \mu \delta_{\alpha \beta} \\
\end{array}
\right) .
\end{eqnarray}


\section{Singlet SC state vs. triplet SC state}
\subsection{Band Calculation and Model Setting}
In the present study temperature is fixed at T = 0.01 [eV].
We take 24 $\times$ 24 meshes for calculating the bands, the susceptibilities, and the superconductivity.
The energy bands are shown in Fig. \ref{fig:bands}(a).
Since 
 the electrons are $3/4$-filled,  
  the Fermi surfaces exist in upper two bands.
The density of states (DOS) is shown in Fig. \ref{fig:fermis}.
There is a large Van Hove singularity above the Fermi energy.

Now we consider the electron correlation effects.
It is expected that $U$ plays important roles for superconductivity in ${\rm (TMTSF)_2ClO_4}$, because
${\rm (TMTSF)_2ClO_4}$ is located next to the spin density wave (SDW) state in the phase diagram 
suggested by Jerome.\cite{jero}
Then we calculate the possible SC states in ${\rm (TMTSF)_2ClO_4}$ 
by choosing $U=0.60$ which is close to the critical value for the divergence of the spin fluctuation in the RPA. 
We investigate the roles of the charge fluctuation by varying the interchain repulsive interactions, $V_a$ 
and $V_b$, with the fixed $U$, where we define $V$ as 
\begin{eqnarray}\label{eq:vdef}
V  \equiv \sqrt{V_a^2 + V_b^2}.
\end{eqnarray}
\begin{figure}[tb]
 \begin{center}
 \includegraphics*[scale=0.8,clip]{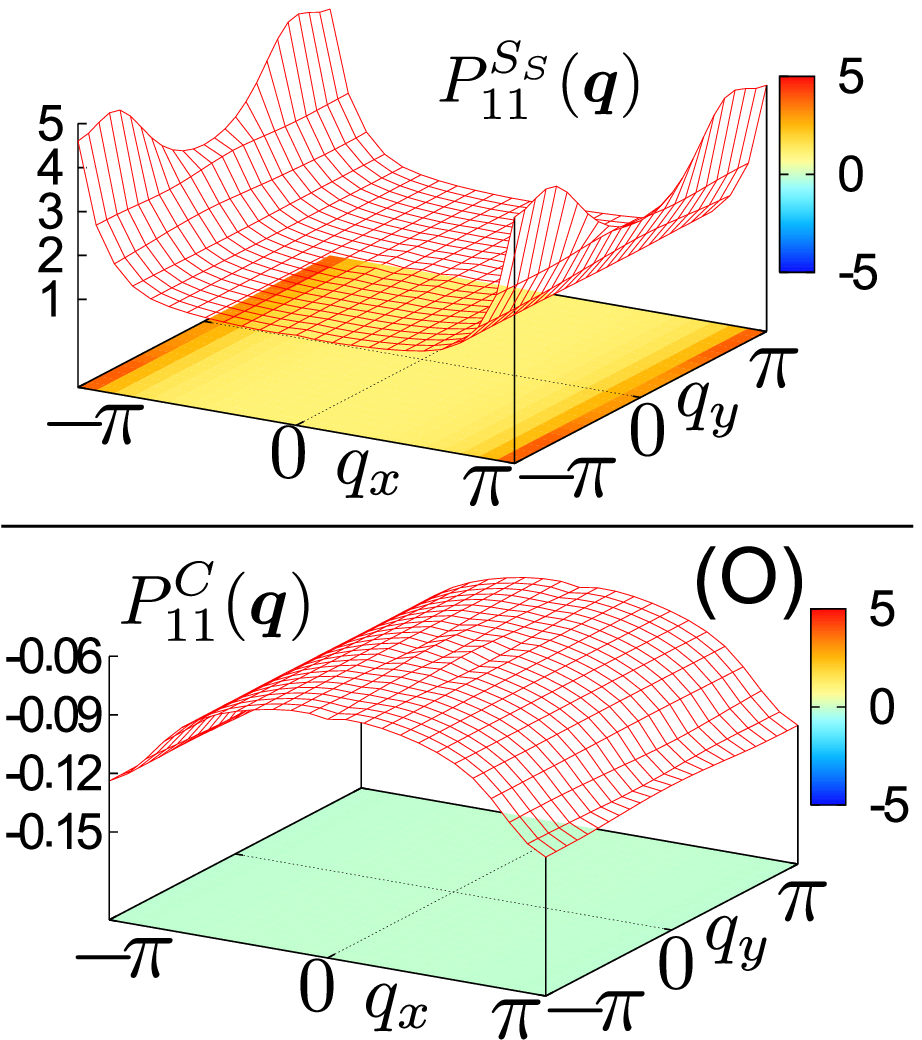}
  \caption{
  The momentum dependence of pairing interactions for the elements $\hat{P}^S_{11}$ 
    and $\hat{P}^C_{11}$ at $T=0.01$ and $V=0$.
  The red and blue represent the repulsive and 
 attractive interactions, respectively. The symbol O corresponds to $V_a=V_b=0$.
  }
 \label{fig:pair}
 \end{center}
\end{figure}

\subsection{Superconductivity for $V$=0}
For $V_a=V_b=0$, 
spin and charge pairing interactions ($\hat{P}^S$ and $\hat{P}^C$) are 
 shown in Fig. \ref{fig:pair}
 where $\hat{P}^S$ is much larger than $\hat{P}^C$.
The susceptibility $\hat{\chi}^S$ which determines $\hat{P}^S$ is strongly 
 enhanced by $U$ compared with  $\hat{\chi}^C$,
 since the  ${\bf q}$ vector  for the peak of $\hat{\chi}^S$ is the same as 
   the nesting vector.

Now we examine numerically Eq. (\ref{eq:linear-gapeq}) 
  to find 
    that $\lambda$ for spin singlet is larger than that for spin triplet, 
 e.g., 
       $\lambda = 0.9$ for the singlet state and 0.3 for the triplet state 
    at $U=0.6$ and $V_a=V_b=0$.   
In the present paper, a phase factor in anomalous self energy is determined 
   in order to maximize  the value $\sum_{\alpha \beta {\bf k}}|{\rm Re}\Sigma^a_{\alpha \beta}({\bf k})|$.
The anomalous self energy consists of 16 elements of  $(\alpha, \beta)$ 
   on the basis of site representation.
Diagonal elements $(\alpha, \alpha)$ 
 are larger than others.
The momentum dependence of these elements is given by 
 "$\cos k_x$" which indicates a sign change on a line separating 
  two Fermi surfaces in Fig. \ref{fig:bands}(b).  
The absence of the node on the Fermi surface, 
 which  gives  a full gap state ( written
  as "singlet (full)"), 
  is consistent with that of Shimahara's result. \cite{shima}
 As for the off-diagonal element of  $\Sigma^a_{\alpha \beta}({\bf k})$, 
 there is a clear contribution  from the same chain ,
  e.g.,  (1,2) and  (3,4), 
   while  that from the interchain is negligibly small.
These behavior of intrachain dependence suggests a role of dimerization, which 
reduces to the half-filled band as shown in Fig. \ref{fig:bands}.   


\begin{figure}[tb]
 \begin{center}
 \includegraphics*[scale=0.3,clip]{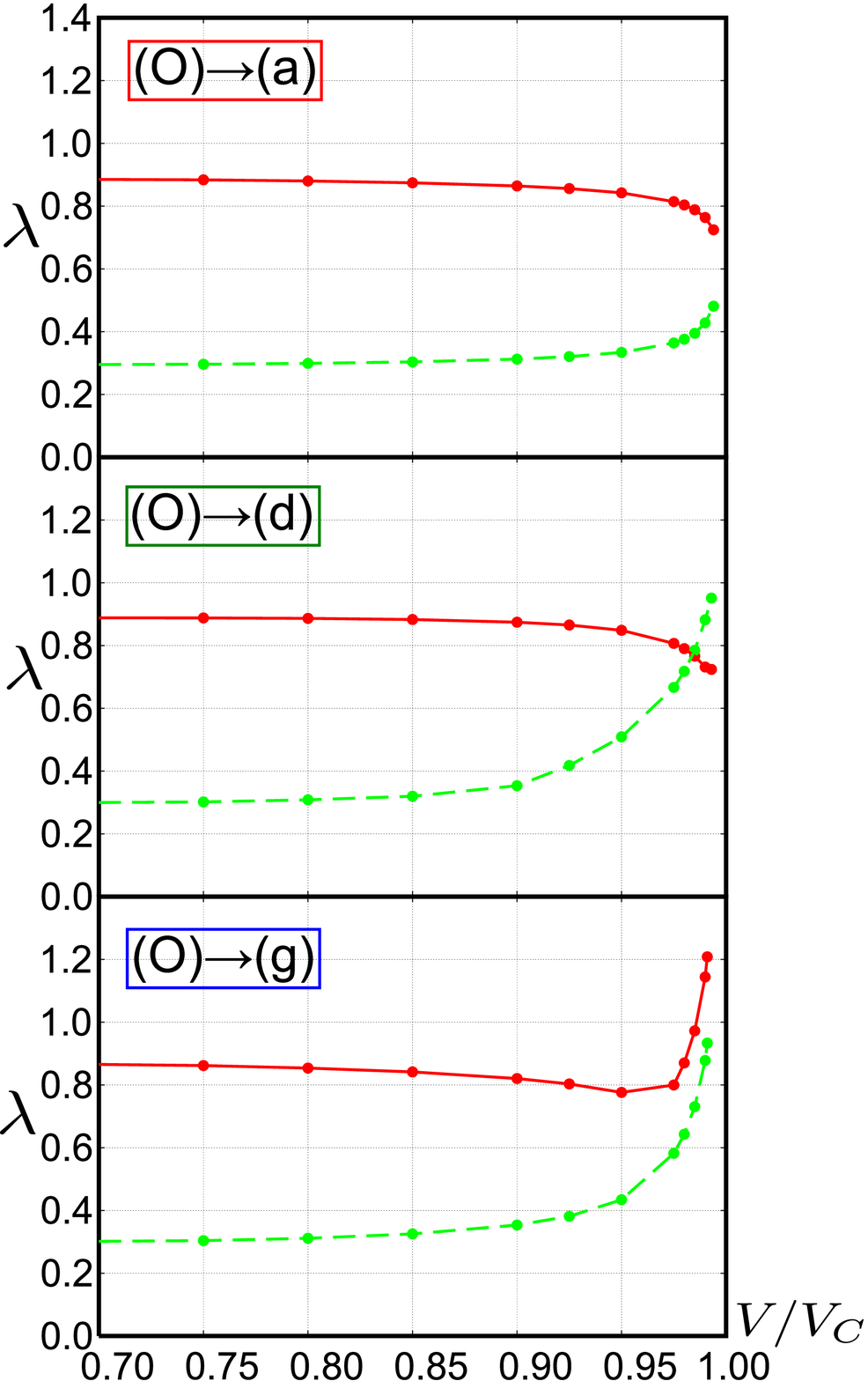}
 \caption{
 V dependence of $\lambda$ for singlet and triplet states for
  $T=0.01, U=0.6$. 
The horizontal axis is shown for $V/V_C > 0.7$ 
   which represents
 from the point (O) to (a), 
 from the point (O) to (d) and  
 from the point (O) to (g), 
 where these points in the parameter space are given by $(V_a,V_b)$=(0,0) (O), (0.00,0.64) (a),  
 (0.33,0.62) (d), and (0.70,0.00) (g), respectively.
  The solid line denotes $\lambda$ for the singlet state 
 and the dashed line denotes $\lambda$ for the triplet state.
 }
 \label{fig:lv}
 \end{center}
\end{figure}
\begin{figure}[tb]
 \begin{center}
 \includegraphics*[scale=0.5,clip]{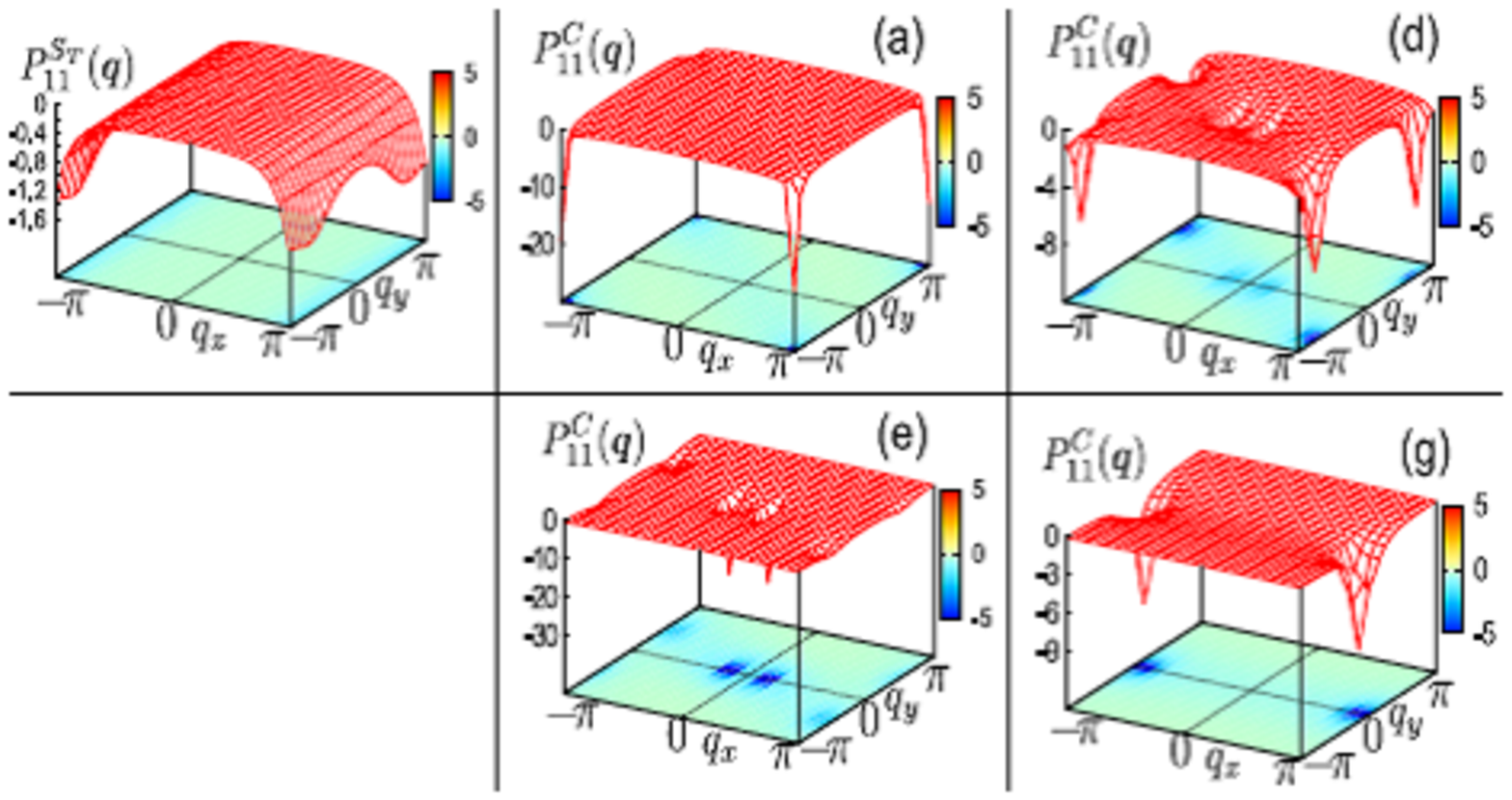}
  \caption{
  The momentum dependence of 
charge-fluctuation induced  pairing interaction 
   ($\hat{P}^C_{11}$) for $V \neq 0$.
where parameters corresponding to points 
(a), (d), (e) and (g) are  (0.00,0.64), (0.33,0.62),
 (0.49,0.49) , and  (0.70,0.00) , respectively.
   The attractive interaction is represented by  blue on the $k_x$-$k_y$ plane.
   For the comparison, $P^{S_T}_{11}$ is shown.
  }
 \label{fig:pair2}
 \end{center}
\end{figure}

\subsection{Superconductivity for $V \neq 0$}
We examine the effect of inter chain interactions 
$V$    
 to find a possibility of  new SC state which is induced  by charge fluctuation. With increasing $V$,  
$\hat{\chi}^C$  increases, and  diverges at  $V=V_C$.  
 In order to calculate the region where 
  the charge fluctuation becomes a dominant one, 
    $V$ is chosen to satisfy the condition 
     that   $\hat{\chi}^C > 3 \hat{\chi}^S$
      and $V < V_C$. 
We examine the SC state on the line from $(V_a,V_b)$=(0,0) (O)  to 
several points given by 
$(V_a,V_b)$=(0.00,0.64) (a), (0.11,0.64) (b),
 (0.21,0.64) (c), (0.33,0.62) (d),
 (0.49,0.49) (e), (0.67,0.22) (f) and
 (0.70,0.00) (g), respectively.
 The ratio of $V_a$ to $V_b$  increases  as the point moves from (a) to (g).
 By solving Eq. (\ref{eq:linear-gapeq}),
   $\lambda$  for singlet and triplet states are calculated and 
   the larger one is chosen as the expected state.

In the region of small $V$, 
 the singlet SC state similar to one at $V=0$ is realized 
  where the state with a full gap is found. 
In  Fig. \ref{fig:lv}, the $V$ dependence of 
  $\lambda$ of both singlet and triplet states is shown  as a function of $V$ 
 for three cases of 
 (O) $\rightarrow$ (a), 
 (O) $\rightarrow$ (d), and  
 (O) $\rightarrow$ (g).  
 For the variation from (O) to  (a), 
 (the first case) , 
    the singlet state is suppressed and 
     the   triplet state is expected.
 For the variation of $V$ from (O) to (d) (the second case), 
 the  singlet state moves to the triplet state close to $V = V_C$.
 For the variation of $V$ from (O) to (g) (the third case), 
 the  singlet state remains  as the dominant state 
  but  moves to another singlet state (as explained later)  
 after  $\lambda$  takes a minimum close to (g). 

Here we note the V dependence of $\lambda$ for triplet state, which 
 is enhanced by  $\hat{\chi}^C$.
 From Eq. (\ref{eq:p-triplet}),
 the term for the spin fluctuation and that for the charge fluctuation has the same sign 
 in triplet pairing interaction.
 Although the triplet SC state is enhanced but the singlet state is  suppressed 
 by charge fluctuations, such an effect becomes large only close to 
 $V = V_C$. This is why the singlet state is favored and the  region 
 for the triplet state is small even in the presence of $V$.  

\section{Pairing interaction and Anomalous self-energy}
Now we examine the respective SC state precisely
 based on the pairing interaction shown in Figs. \ref{fig:pair2} and anomalous self-energy.

(i) First, we discuss the case indicating a possible triplet SC state, which 
 is given by (a) and (d).
 When  $\hat{\chi}^C$ is increased by $V$,
 the singlet state is  suppressed but the triplet state is enhanced 
 (Fig. \ref{fig:pair2} (a) and (d)). 
   Actually, the competition between these two kinds of fluctuations occurs 
      for the  singlet state as seen from  Eq. (\ref{eq:p-singlet}) and 
 by noting a fact that  
  the peak position of  $\hat{P}^C$  is nearly the same as that of 
     $\hat{P}^S_S$ 
      from Fig. \ref{fig:pair}.  
  The triplet state may be expected when   $V \rightarrow V_C$,  
    and is rather favored by the introduction of $V_a$. 
 When the point moves from (a) to (d) (i.e., increasing $V_a$),  
   the peak position of $\hat{P}^C$ begins to deviate slightly from $(\pi,\pi)$
   and the attractive interaction increases due to 
     the increase of the amount of $\hat{P}^C$ integrated with respect 
      to the momentum.

(ii) Secondly, we study  the region
     where the singlet state is favored  by the effects 
         of $\hat{\chi}^C$ (points (e) - (g).
 The pairing interactions ($\hat{P}^C$) 
     induced by charge fluctuation  
          are shown in Figs. \ref{fig:pair2} (e) and (g)  
        where the position of the peak   is 
      different from that of in $\hat{\chi}^S$.
As a result, the effect of $\hat{\chi}^C$ on the singlet pairing state
    becomes  compatible with that of $\hat{\chi}^S$ 
       for  $V \rightarrow V_c$.
Further 
since $\hat{P}^C$ is larger than $\hat{P}^S$,  
 the  spin singlet state is determined mainly by 
    $\hat{\chi}^C$ and is assisted by   $\hat{\chi}^S$.
 The change of the self-energy  along $q_x$ axis is very small. 
 This comes from the difference in 
   the peak position of charge fluctuations i.e., 
  $(\pi,\pi)$ for (a) and that on the $q_x$ axis for 
        (e) and (g). 



\begin{figure}[tb]
 \begin{center}
 \includegraphics*[scale=0.8,clip]{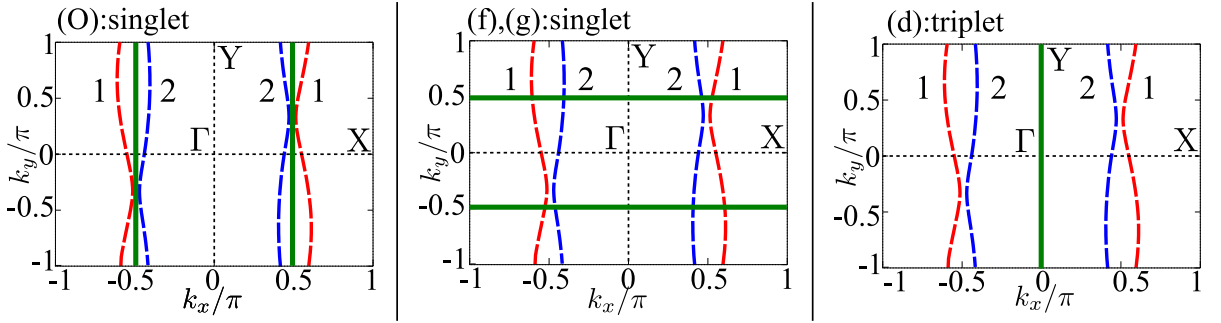}
 \caption{
 Fermi surface and gap node where  
  the dashed lines denotes Fermi surface and the solid line
    denote gap node  at $T=0.01,U=0.6$.
 }
 \label{fig:nodes}
 \end{center}
\end{figure}

Finally, we examine the node of the gap, which is calculated from Eqs. (\ref{eq:bcsH}) and 
(\ref{eq:diago2}).
Figure \ref{fig:nodes} denotes the position of nodes and Fermi surfaces 
   for points (O), (f), (g) and (d).
 For the  singlet state (O), which is mediated by $\hat{\chi}^S$ 
   the node exists between two different Fermi surfaces
  implying a full gap due to the existence of the  two bands.\cite{shima}
 For (f) and (g), 
 the singlet state mediated by $\hat{\chi}^C$ 
   has two nodes at  $k_y = \pm \pi/2$, 
     which cross the  Fermi surfaces.
Gap function for triplet has a node on the line $k_x=0$ suggesting a full-gap state.

\section{Summary and Discussion}

We have found the possible singlet SC state with the line nodes 
in the extended multi-site Hubbard model 
representing ${\rm (TMTSF)_2ClO_4}$ with the multi-band and under the anion ordering.
Such a state is induced by the charge fluctuation
 owing to the interchain Coulomb interaction $V_a$ (along the $y$-axis), 
while the singlet SC state with anisotropic (but finite) gap suggested by Shimahara\cite{shima}
 is reproduced in the presence of the strong spin fluctuation.
Moreover, we have investigated the roles of the charge fluctuation for the triplet SC state.
It is found that the triplet state appears in a certain range 
of the interchain Coulomb interactions, $V_a$ and $V_b$, 
It is noted that the charge fluctuation cooperates with the spin fluctuation for the triplet pairing, 
but competes for the singlet pairing. 
For such a triplet state, the p-wave like state is obtained 
  in the present multi-site model, 
 while the f-wave gap function have been suggested in the single site models.
\cite{tanaka,kuroki,aizawa,aizawa2,bel,duprat,fuseya,nick}

Finally we discuss how to distinguish the SC state with the lines and that with full gap of anisotropy on the Fermi surface.
Using a maximum, $\Delta_{\rm max}$, and a minimum, $\Delta_{\rm min}$,
for the latter case, $1/T_1$ reveals the exponential behavior at low temperature, i.e.,
 $T < \Delta_{\rm min}$.
In the preset case, it is expected that $1/T_1$ exhibits exponential decrease at lower temperatures than 
$T_{\rm c} \Delta_{\rm min} / \Delta_{\rm max} \sim 0.2$K using the ratio given by both our and 
Shimahara's results.
On the other hand, Yonezawa et al. have reported that the $1/T_1$ exhibits $T^3$ behavior 
at temperatures lower than $0.2$K\cite{yone}.
This fact suggests that the SC state with the line nodes supports the result of the NMR experiment 
in the absence of magnetic field.

\section*{Acknowledgment}
We are thank to S. Yonezawa, Y. Maeno, H. Shimahara for useful discussions.
This work was financially 
supported in part by a Grant-in-Aid for Special Coordination
Funds for Promoting Science and Technology (SCF),
Scientific Research on Innovative Areas 20110002, and
Scientific Research 19740205 from the Ministry of Education,
Culture, Sports, Science and Technology of Japan.







\begin{thebibliography}{99}
\bibitem{Bechgaard} K. Bechgaard, K. Carneiro, M. Olsen, F. B. Rasmussen, and C. S. Jacobsen 1981 {\it Phys. Rev. Lett.} \textbf{46} 852.
\bibitem{taki} M. Takigawa, H. Yasuoka, and G. Saito 1987 {\it J. Phys. Soc. Jpn.} \textbf{56} 873-876.
\bibitem{jero} D. J\'erome 1991 {\it Sciense} \textbf{252} 1509.
\bibitem{shina} J. Shinagawa, Y. Kurosaki, F. Zhang, S. E. Brown, D. J\'erome, J. B. Christensen, and K. Bechgaard 2007 {\it Phys. Rev. Lett.} \textbf{98} 147002.
\bibitem{lee} I. J. Lee, M. J. Naughton, G. M. Danner and P. M. Chaikin 1997 {\it Phys. Rev. Lett.} \textbf{78} 3555.
\bibitem{lee2} I. J. Lee, S. E. Brown, W. G. Clark, M. J. Strouse, M. J. Naughton, W. Kang, and P. M. Chaikin 2002 {\it Phys. Rev. Lett.} \textbf{88} 017004.
\bibitem{lee3} I. J. Lee, D. S. Chow, W. G. Clark, M. J. Strouse, M. J. Naughton, P. M. Chaikin, and S. E. Brown 2003 {\it Phys. Rev. B} \textbf{68} 092510.
\bibitem{yone} S. Yonezawa, S. Kusaba, Y. Maeno, P. Auban-Senzier, C. Pasquier, K. Bechgaard, and D. J\'erome 2008 {\it Phys. Rev. Lett.} \textbf{100} 117002.
\bibitem{yone2} S. Yonezawa, S. Kusaba, Y. Maeno, P. Auban-Senzier, C. Pasquier, and D. J\'erome 2008 {\it J. Phys. Soc. Jpn.} \textbf{77} 054712.
\bibitem{yone3} S. Kusaba, S. Yonezawa, Y. Maeno, P. Auban-Senzier, C. Pasquier, K. Bechgaard, D. J\'erome 2008 Solid State Science {\bf 10} 1768-1772.
\bibitem{yoshi} H. Yoshino, A. Oda, T. Sakaki, T. Hanajiri, J. Yamada, S. Nakatsuji, H. Anzai and K. Murata 1999 {\it J. Phys. Soc. Jpn.} \textbf{68} 3142.
\bibitem{pou} J. P. Pouget, G. Shirane, K. Bechgaard and J. M. Fabre 1983 {\it Phys. Rev. B} \textbf{27} 5203.
\bibitem{band} D. Le P\'evelen, J. Gaultier, Y. Barrans, D. Chasseau, F. Castet, and L. Ducasse 2001 {\it The European Physical Journal B} \textbf{19} 363.
\bibitem{tanaka}  Y. Tanaka, and K. Kuroki 2004 {\it Phys. Rev. B} \textbf{70} 060502.
\bibitem{kuroki} K. Kuroki, and Y. Tanaka 2005 {\it J. Phys. Soc. Jpn.} \textbf{74} 1694.
\bibitem{aizawa}  H. Aizawa, K. Kuroki, and Y. Tanaka 2008 {\it Phys. Rev. B} \textbf{77} 144513.
\bibitem{aizawa2} H. Aizawa, K. Kuroki, T. Yokoyama, and Y. Tanaka 2009 {\it Phys. Rev. Lett.} \textbf{102} 016403.
\bibitem{bel} N. Belmechri, G. Abramovici, and M. Heritier 2008 {\it Eur. Phys. Lett.} \textbf{82} 47009.
\bibitem{kinokon} H. Kino, and H. Kontani 1999 {\it J. Phys. Soc. Jpn.} \textbf{68} 1481.
\bibitem{duprat} R. Duprat, and C. Bourbonnais 2001 {\it Eur. Phys. J. B.} \textbf{21} 219.
\bibitem{fuseya} Y. Fuseya, and Y. Suzumura 2005 {\it J. Phys. Soc. Jpn.} \textbf{74} 1263.
\bibitem{nick} J. C. Nickel, R. Duprat, C. Bourbonnais, and N. Dupuis 2005 {\it Phys. Rev. Lett.} \textbf{95} 247001.
\bibitem{shima} H. Shimahara 2000 {\it Phys. Rev. B} \textbf{61} 14936.
\bibitem{hase} Y. Hasegawa and K. Kishigi 2008 {\it Phys. Rev. B} \textbf{78} 045117.
\bibitem{koba} A. Kobayashi, Y. Tanaka, M. Ogata, and Y. Suzumura 2004 {\it J. Phys. Soc. Jpn.} \textbf{73} 1115.
\bibitem{koba2} A. Kobayashi, S. Katayama, K. Noguchi, and Y. Suzumura 2004 {\it J. Phys. Soc. Jpn.} \textbf{73} 3135.
\end{thebibliography}








\end{document}